\documentclass{ifacconf}

\usepackage{graphicx}      % include this line if your document contains figures
\usepackage{natbib}        % required for bibliography

\graphicspath{Figures/}
\DeclareGraphicsExtensions{.png,PNG,.pdf,.PDF}
\usepackage{bbm}
\usepackage{bm}
\usepackage{amsmath} % assumes amsmath package installed
\protected\edef\ell{\noexpand\ensuremath{{\mathchar\the\ell}}}
\usepackage{amssymb}  % assumes amsmath package installed
\usepackage{mathrsfs}
\usepackage{multicol}
\usepackage{xcolor}
\usepackage{subcaption}
\usepackage[ruled,linesnumbered]{algorithm2e}
\usepackage{mathtools}
\usepackage{soul}

\allowdisplaybreaks
\usepackage{eucal}
\usepackage{url}
\DeclareMathAlphabet{\pazocal}{OMS}{zplm}{m}{n}

\newcommand{\cC}{\mathbb{C}}
\newcommand{\cG}{\pazocal{G}}
\newcommand{\cE}{\pazocal{E}}
\newcommand{\cV}{\pazocal{V}}
\newcommand{\until}[1]{\{1,\dots, #1\}}

\begin{document}
\begin{frontmatter}

\title{Multi-agent Distributed Model Predictive Control with Connectivity Constraint} 

\thanks{This research has been supported by the Italian Ministry of University and Research (MIUR) under the PRIN 2017 grant n. 201732RS94 ``Systems of Tethered Multicopters''.
}%

\author[First]{Andrea Carron} 
\author[Second]{Danilo Saccani} 
\author[Second]{Lorenzo Fagiano}
\author[First]{Melanie N. Zeilinger} 

\address[First]{Institute for Dynamic Systems and Control, ETH Zurich, Switzerland. (email \{carrona, mzeilinger\}@ethz.ch)}
\address[Second]{Dipartimento di Elettronica, Informazione e Bioingegneria, Politecnico di Milano, Piazza Leonardo da Vinci 32, Milano, Italy. (email: \{danilo.saccani, lorenzo.fagiano\}@polimi.it) }

\begin{abstract}                % Abstract of not more than 250 words.
In cooperative multi-agent robotic systems, coordination is necessary in order to complete a given task. Important examples include search and rescue, operations in hazardous environments, and environmental monitoring. Coordination, in turn, requires simultaneous satisfaction of safety critical constraints, in the form of state and input constraints, and a connectivity constraint, in order to ensure that at every time instant there exists a communication path between every pair of agents in the network. In this work, we present a model predictive controller that tackles the problem of performing multi-agent coordination while simultaneously satisfying safety critical and connectivity constraints. The former is formulated in the form of state and input constraints and the latter as a  constraint on the second smallest eigenvalue of the associated communication graph Laplacian matrix, also known as Fiedler eigenvalue, which enforces the connectivity of the communication network. We propose a sequential quadratic programming formulation to solve the associated optimization problem that is amenable to distributed optimization, making the proposed solution suitable for control of multi-agent robotics systems relying on local computation. Finally, the effectiveness of the algorithm is highlighted with a numerical simulation.
\end{abstract}


\end{frontmatter}
%===============================================================================
\section{Introduction}
In the last two decades, many applications employed cooperative multi-agent setups, for example, environmental monitoring, search and rescue operations and exploration of unknown environments.
%In the last two decades, cooperative multi-agent setups were employed in many different applications, for example, environmental monitoring, search and rescue operations, and exploration of unknown environments, see~\cite{SINOPOLI:2003,THURN:2005,BAXTER:2007,QUERALATA:2020}.
%In the last two decades, many researchers focused their attention on the coordination of multi-agent systems, and this topic is still very active due to its many challenges and increasing on-board communication/computation capabilities.
Compared to single-agent setups, multi-robot systems can solve certain tasks faster and more robustly, moreover some tasks can just not be solved by a single agent. 
%Applications span from environmental monitoring and search and rescue operations, to exploration of unknown environments, see for example~\cite{SINOPOLI:2003,THURN:2005,BAXTER:2007,QUERALATA:2020}.
However, to successfully deploy a multi-agent system, it is necessary to simultaneously guarantee safety and communication network connectivity,~\cite{ROBUFFO_CONNECTIVITY:2013}. The former is commonly formulated in the form of satisfaction of safety critical state constraints, like collision avoidance with obstacles and other agents. The latter requires the existence of a multi-hop communication path between each pair of agents in the communication network. In most distributed multi-agent problems, connectivity is usually assumed rather than enforced, see for example~\cite{JADBABAIE_COORDINATION:2003,OLFATI_CONSENSUS:2004,CONTE_DMPC:2016, CARRON:2021}. However, the loss of connectivity can lead to the failure of the controller as information cannot propagate throughout the whole network. Guaranteeing connectivity in multi-robot systems is particularly challenging due to the time-varying nature of the communication network that depends on the agents relative positions. 

\emph{Contributions:} In this work, we propose a multi-agent model predictive controller (MPC) for tracking that can simultaneously guarantee satisfaction of safety critical and connectivity constraints. The latter is enforced by imposing a constraint on the second smallest eigenvalue of the Laplacian matrix. The proposed controller maintains the MPC theoretical results of closed-loop constraint satisfaction and convergence. Moreover, we propose a Sequential Quadratic Programming (SQP) reformulation of the original nonlinear optimal control problem that is amenable to distributed implementation, i.e., it relies on local communication.

\emph{Related Work:}
In the last decades, model predictive control received a considerable attention thanks to its capability of systematically handling state and input constraints.  The application of MPC to control multi-agent and distributed systems has also been widely studied, see for example~\cite{scattolini2009architectures} and~\cite{negenborn2009multi}. While there is a variety of techniques, the approach most relevant to this paper is the work presented by~\cite{FERRAMOSCA:2009} and~\cite{limon2018nonlinear}, where the authors present a tracking MPC formulation for nonlinear constrained systems.

\cite{SPANOS_CONNECTIVITY:2004} were among the first to consider connectivity in a control problem. The work introduces the concept of geometric robust connectivity, which however, in the examples presented induces unwanted chattering behavior of the agents' inputs. Alternative approaches for formation control have been presented in~\cite{FAX_FORMATION:2004,NOTARSTEFANO_CONNECTIVITY:2006}, but they enforce the maintenance of local connections, which can be very restrictive. An alternative approach, based on enforcing connectivity on the adjacency matrix, is introduced in~\cite{ZAVLANOS_CONNECTIVITY:2005}, however, the resulting controller cannot be computed using only local information. The successful idea of using the Fiedler eigenvalue, see~\cite{FIEDLER_CONNECTIVITY:1989}, to preserve connectivity is introduced in~\cite{DEGENNARIO_CONNECTIVITY:2006,ZAVLANOS_CONNECTIVITY:2007}. In~\cite{DEGENNARIO_CONNECTIVITY:2006}, the authors make direct use of the Fiedler eigenvalue, while in~\cite{ZAVLANOS_CONNECTIVITY:2007}, the authors exploit the determinant of the reduced Laplacian. Applications of the two proposed methods to the problem of formation control and rendezvous problems, as well as the integration of radio signal strength maps are considered in~\cite{HSIEH_SIGNAL_MAP:2008,JI_CONNECTIVITY:2007}. In~\cite{YANG_CONNECTIVITY:2010}, a more efficient distributed method for computing the Fiedler eigenvalue is presented, and in~\cite{SABATTINI_CONNECTIVITY:2012,ROBUFFO_CONNECTIVITY:2013}, the authors introduce two connectivity maintenance algorithms for Lagrangian and port-Hamiltonian non-linear systems, respectively. Finally, we refer the readers to~\cite{ZAVLANOS_UNIFED_CONNECTIVITY:2011}  for a unified framework for connectivity maintenance. To the best of authors' knowledge, this is the first work where connectivity is imposed as constraint using an optimization-based control framework. Moreover, while some of the aforementioned approaches provide heuristics to satisfy constraints, to the best of the authors' knowledge none of the previous results provide formal guarantees.

The remainder of the paper is organized as follows. In Section~\ref{sec:problem_formulation}, we introduce the problem setup. Section~\ref{sec:method} and~\ref{sec:sqp} describe the proposed method and an SQP-based solution for the resulting optimization problem, respectively. In Section~\ref{sec:numerical}, we validate the controller with numerical simulations. Finally, Section~\ref{sec:conclusions} concludes the paper. 

%%%%%%%%%%%%%%%%%%%%%%%%%%%%%%%%%%%%%%%
\section{Problem Formulation}
\label{sec:problem_formulation}
This section introduces the agents dynamics and their constraints, as well as the communication assumptions. Finally, we formally introduce the problem we aim to solve.
\subsection{Agents Dynamics and Constraints}
We consider a swarm of $M$ agents, each of them modeled as a non-linear time-invariant dynamical system of the form
\begin{equation}
    \label{eq:model}
    \begin{split}
        &x_i(k+1) = f_i(x_i(k),u_i(k)), \\
        &p_i(k) = C_ix_i(k),
    \end{split}
\end{equation}
where $x_i \in \mathbb{R}^{n_i}$, $u_i \in \mathbb{R}^{m_i}$, and $p_i \in \mathbb{R}^{o_i}$ are the state, input, and position of agent~$i$, respectively. In most robotics applications, the position vector is either 2 or 3-dimensional, i.e., $o_i = 2$ or $3$. Let us denote with $(\bar{x}_i, \ \bar{u}_i)$ an equilibrium of system~\eqref{eq:model}, and we assume that each agent is able to measure its own state~$x_i$.
Let us consider, without loss of generality, that the position of the system $p_i$ is at the top of the state vector~$x_i$ and introduce the operator~$\psi (p_i) = [p_i^T,0, \dots, 0]^T\in\mathbb{R}^{n_i}$ that generates a vector of the dimension of $x_i$ with first entry, corresponding to the position $p_i$ of the $i$-th robot.
\begin{assum}
\label{ass:position_invariance}
The agents' dynamics~\eqref{eq:model} are position invariant, i.e. $\forall p_j \in \mathbb{R}^{o_i},\ x_i,\ u_i, \ f_i(x_i+\psi(p_j),u_i)=x_i(k+1)+\psi(p_j)$.
\end{assum}
Assumption \ref{ass:position_invariance} is in particular satisfied when the position is the output of an integrator and no forces or moments depend on it, which is a standard condition in many ground and aerial robots. The state and input variables are constrained to lie within the following sets
\begin{equation}
    \label{eq:constraints}
    \begin{split}
        &x_i \in \mathbb{X}_i,\ \ \ u_i \in \mathbb{U}_i.
    \end{split}
\end{equation}
Finally, by stacking the state of each agent in a vector and similarly for the inputs, we obtain the swarm state~$x=\text{col}_{i\in\until{M}} x_i$ and input~$u=\text{col}_{i\in\until{M}} u_i$.

\subsection{Communication Network and Generalized Connectivity}
In the proposed setup, agent $i$ can communicate  with spatially close neighbours, i.e., all agents contained within a set $\cC_i (p_i)$, centered at the agent's position~$p_i: \ \psi(p_i) \in \mathbb{X}_i$. For example, assume that agents $i$ and $j$ have a communication set $\cC_l(p_l)=\{q:\|p_l - q\|_2 \leq d \}$, with~$l\in\{i,j\}$ and $d > 0$ being their maximum communication range. Thus, the agents are able to communicate if and only if $\|p_i - p_j\|_2 \leq d$.
The communication network is defined as an undirected and time-varying communication graph~$\cG(k)=(\cV, \cE(k))$, where the set of nodes~$\cV\in\until{M}$ represents the agents in the network, and the set of edges~$\cE(k) \subset\cV \times \cV$ contains the pairs of agents~$\{i,j\}$, which can communicate with each other at time~$k$. $\cE(k)$ can change over time as the edges depend on the agents' relative position. The set of neighbors of node~$i$, including~$i$ itself, is denoted by $\pazocal{N}_i = \{i\} \cup \{j\ |\ \{i,j\}\in\cE(k)\} \subseteq \cV$. The state of the agent~$i$  with the state of its neighbours is denoted $x_{\pazocal{N}_i}$. To include possible interactions between neighbouring subsystems we consider also the following coupling state constraint: 
\begin{equation}\label{eq:couplingConstr}
    c(x_{\pazocal{N}_i}) \leq 0.
\end{equation}
Constraint~\eqref{eq:couplingConstr} is left purposely general since it can represent any coupling between neighbouring subsystems. A typical example in multi-robot systems is collision avoidance between neighbours.

The Laplacian matrix~$L\in\mathbb{R}^{M \times M}$ of a graph~$\cG$ is defined as~$L = D - A$, where $A$ is the adjacency matrix and $D$ is the degree matrix of the graph. The adjacency matrix~$A \in \mathbb{R}^{M \times M}$ is a square symmetric matrix with elements~$A_{ij} \geq 0$ such that~$A_{ij} = 0$ if $\{i, j\} \notin \cE(k)$, $A_{ii} = 0$, and $A_{ij} > $0 otherwise. The matrix~$D \in \mathbb{R}^{M \times M}$ is a diagonal matrix where~$D_{ii} = \sum_{j=1}^M A_{ij}$.
A graph is said to be connected if there is a sequence of edges between any pair of vertices. A measure for the graph connectivity is given by the second smallest eigenvalue~$\lambda_2$ of the Laplacian matrix~$L$ of the graph~$\cG$, also called the Fiedler eigenvalue or connectivity eigenvalue, see for example~\cite{BULLO_NETWORKS:2022,oellermann1991laplacian}. When the graph~$\cG$ is connected, the connectivity eigenvalue~$\lambda_2$ is strictly positive, i.e.,
\begin{equation}
    \label{eq:connectivity}
    \lambda_2(L) > 0.
\end{equation} 
Making the adjacency matrix~$A$ dependent on the agents' state $x_i$ may lead to a loss of connectivity in specific scenarios. We define a smooth function~$\phi(x_i,x_j)$ such that~$\phi(x_i,x_j) > 0$ if and only if~$\|p_i - p_j\|_2 \leq d$ and~$\phi(x_i,x_j) = 0$ otherwise. The function~$\phi(x_i,x_j)$ is used to weigh the inter-agent links in the adjacency matrix, i.e., by setting~$A_{ij} = \phi(x_i,x_j)$. The connectivity condition~\eqref{eq:connectivity} thus becomes state dependent
\begin{equation}
    \label{eq:generalized_connectivity}
    \lambda_2(L(x)) > 0.
\end{equation}
Condition~\eqref{eq:generalized_connectivity} is often referred to as generalized connectivity, see~\cite{ROBUFFO_CONNECTIVITY:2013}. 

In the next sections, we will make use of the following result, providing that both eigenvalues and eigenvectors of the Laplacian matrix~$L$ are differentiable under the following assumption:
\begin{assum} \label{ass:secondeig}
The second smallest eigenvalue $\lambda_2(L(x))$ of the graph Laplacian matrix is simple $\forall k\geq 0$.
\end{assum}
\begin{thm}{(Theorem 1 in \cite{magnus1985differentiating})} \label{th:theremEig}
    Let Assumption \ref{ass:secondeig} hold, and let $X_0$ be a real symmetric $n\times n$ matrix. Let $v_0$ be a normalized eigenvector associated with a simple eigenvalue $\lambda_0$ of $X_0$. Then a real-valued function $\lambda$ and a vector function $v$ exist for all $X$ in some neighbourhood $\mathcal{N}(X_0)\in\mathbb{R}^{n\times n}$ of $X_0$, such that:
    \begin{equation}
        \lambda(X_0)=\lambda_0, \ \ \ \ v(X_0)=v_0,
    \end{equation}
    and
    \begin{equation}
        Xv=\lambda v,  \ \ \ \ v^Tv=1, \ \ \ \ X\in \mathcal{N}(X_0).
    \end{equation}
    Moreover, the functions $\lambda$ and $v$ are $\infty$ times differentiable on $\mathcal{N}(X_0)$, and the differentials at $X_0$ are:
    \begin{equation}
        \text{d}\lambda = v_0^T(\text{d}X)v_0 \label{eq:dsigma}
    \end{equation}
    and
    \begin{equation}
        \text{d}v=(\lambda_0 I_n-X_0)^+(\text{d}X)v_0 \label{eq:dv}
    \end{equation}
    where $X^+$ denotes the Moore-Penrose inverse of matrix $X$ and $I_n$ denotes the $n$-by-$n$ identity matrix.
\end{thm}
\begin{rem}
In order for $\lambda$ and $v$ to be differentiable at $X_0$ we require that $\lambda_0$ is simple, but this does not exclude the possibility of multiplicities among the remaining $n-1$ eigenvalues of $X_0$. This assumption is usually implicitly considered in approaches for connectivity maintenance, see for example~\cite{YANG_CONNECTIVITY:2010,ROBUFFO_CONNECTIVITY:2013}. In our simulations, this assumption is always verified; however, to formally guarantee the assumption, it can also be enforced through a nonlinear constraint in the proposed FHOCP.
\end{rem}
\subsection{Problem Statement}
Based on the defined setup, the problem addressed in this paper is formally stated in the following. 

\begin{prob} \label{pr:problem}
Consider a swarm of $M$ robots with dynamics~\eqref{eq:model} and subject to constraints~\eqref{eq:constraints},~\eqref{eq:couplingConstr}. The agents' initial condition $x(0) = x_0$ satisfies the generalized connectivity condition $\lambda_2(L(x_0)) > 0$. Given $M$ reference targets $r = [r_1, \ldots, r_M]^T$, such that $\lambda_2\left(L\left([\psi(r_1), \ldots, \psi(r_M)]^T\right)\right) > 0$, determine control inputs $u$ such that generalized connectivity condition~\eqref{eq:generalized_connectivity} and constraints~\eqref{eq:constraints} are satisfied at all times $k>0$ and, if feasible, $\lim \limits_{k\rightarrow\infty} \| p(k) - r\|_2 \rightarrow 0$.
\end{prob}

%%%%%%%%%%%%%%%%%%%%%%%%%%%%%%%%%%%%%%%
\section{MPC with connectivity constraint}
\label{sec:method}
In this section, we introduce a method to solve Problem~\ref{pr:problem} that simultaneously takes into account the non-linear dynamics of the agents~\eqref{eq:model}, state and input constraints~\eqref{eq:constraints} and~\eqref{eq:couplingConstr}, and the connectivity constraint~\eqref{eq:connectivity}. To this aim, we propose a nonlinear tracking MPC inspired by the work in \cite{limon2018nonlinear}, where an artificial reference is considered as an additional decision variable. Problem~\eqref{pr:problem} is cast as the following Finite Horizon Optimal Control Problem (FHOCP), where $x(j|k)$ denotes the open-loop prediction of the variable $x$ at time $j+k$ predicted at time $k$.
\begin{subequations}\label{eq:FHOCP}
    \begin{align}
        &\min \limits_{U,\bar{x},\bar{u},\bar{r}} \;\;
          \sum_{i=1}^{M}{J_i(x,U_i,\bar{x}_i,\bar{u}_i,r_i,\bar{r}_i)} \label{seq:cost}\\\
         &\text{subject to: }\nonumber\\
         & x_{i}(0|k)=x_i(k) \label{seq:initialcondition}\\
         & x_{i}(j+1|k)=f_i(x_{i}(j|k),u_{i}(j|k)) &&\forall j\in\mathbb{N}_0^{N-1} \label{seq:sysdyn}\\
         & p_{i}(j|k)=C_i x_{i}(j|k) &&\forall j\in\mathbb{N}_0^{N} \label{seq:output}\\
         & \bar{x}_i = f_i(\bar{x}_i,\bar{u}_i)\\
         & x_{i}(N|k)=\bar{x}_i  \label{seq:steadyState}  \\
         & \bar{r}_i = C_i \bar{x}_i \\
         & x_{i}(j|k)\in\mathbb{X}_i && \forall j\in\mathbb{N}_0^{N} \label{seq:stateconstraints}\\
         & c(x_{\pazocal{N}_i}(j|k))\leq 0 && \forall j\in\mathbb{N}_0^{N} \label{seq:couplingstateconstraints}\\
         & u_{i}(j|k)\in\mathbb{U}_i && \forall j\in\mathbb{N}_0^{N-1}\label{seq:inputconstraints}\\
         & \lambda_2\left(L(x(j|k))\right) >0 && \forall j\in\mathbb{N}_0^{N} \label{seq:connetivityConstr} \\
         & \forall i\in\mathbb{N}_1^M,
    \end{align}
\end{subequations}
where $U_i$ is the input sequence of agent $i$ over a prediction horizon $N \in \mathbb{N}$, $U=\text{col}_{i\in\{1,\dots,M\}} U_i$ the vector containing all the input sequences, $J_i$ is the cost function of agent $i$, which depends on input, state, reference and artificial reference of agent $i$, and $\mathbb{N}_i^{j}$ indicates the integers from $i$ to $j$. Constraints~\eqref{seq:initialcondition}-\eqref{seq:output}
represent the initial condition and dynamics of the system, constraints~\eqref{seq:stateconstraints}-\eqref{seq:inputconstraints} pertain to state and input,~\eqref{seq:steadyState} is the terminal constraint, and~\eqref{seq:connetivityConstr} the connectivity one. 

Notice that Problem~\eqref{eq:FHOCP} couples the agents via constraint~\eqref{seq:couplingstateconstraints}, which couples neighbouring agents, and connectivity constraint~\eqref{seq:connetivityConstr}, which couples the state of all the agents in the system. Since coupling constraints of the form~\eqref{seq:couplingstateconstraints} have been widely addressed in the distributed MPC literature, see for example~\cite{mueller2017economicdistributed}, we will mainly focus on how to treat the connectivity constraint~\eqref{seq:connetivityConstr} in the following.

The different ingredients defining the FHOCP~\eqref{eq:FHOCP} are described in the following subsections.

\subsubsection{Cost Function}
Given the reference $r_i$ for each agent, the cost function considered in~\eqref{seq:cost} is defined as
\begin{multline} \label{eq:CostFun}
     J_i(x,U,\bar{x}_i,\bar{u}_i,r_i,\bar{r}_i)) = \sum_{j=1}^{N-1} l_i(x_{i,(j|k)}-\bar{x}_i,u_{i,(j|k)}-\bar{u}_i)+ \\ +l_{i,N}(\bar{r}_i-r_i),
\end{multline}
where $l_i(\cdot,\cdot):\mathbb{R}^{n_i}\times\mathbb{R}^{m_i}\rightarrow \mathbb{R}$ is the stage cost function and $l_{i,N}(\cdot):\mathbb{R}^{o_i}\rightarrow \mathbb{R}$ is an offset cost that weighs the distance between the final position of the predicted trajectory, i.e., the artificial reference, and the reference $r_i$. Let us consider the following standard assumption, see~\cite{limon2018nonlinear}, on the stage and offset cost.
\begin{assum} \label{ass:cost}
Consider the cost function~\eqref{eq:CostFun}. The stage cost function $l_i(\cdot,\cdot)$ is positive definite, and the offset cost $l_{i,N}(\cdot)$ is a positive definite, convex subdifferentiable function.
%\begin{itemize}
%    \item Let $l_i(\cdot,\cdot)$ be positive definite functions.
%    \item Let $l_{i,N}(\cdot)$ be positive definite, convex subdifferentiable functions.
%\end{itemize}
\end{assum}

\subsubsection{Connectivity Constraint}
As mentioned in Section~\ref{sec:problem_formulation}, defining the adjacency matrix $A$ via the weight function~$\phi(x_i,x_j)$ renders the connectivity condition state-dependent. We consider continuously differentiable functions~$\phi$, allowing us to compute the Jacobian and Hessian with of the Laplacian $L$ with respect to the state, ultimately needed to solve the optimization problem.
As an example, in \cite{YANG_CONNECTIVITY:2010}, the following position-dependent function has been considered:
\begin{equation}
    \phi(p_i,p_j)=e^{-\|p_i-p_j \|_2^2 / 2\sigma^2},
\end{equation}
where $\sigma$ is a scalar parameter computed such that $e^{-\rho_c^2 / 2\sigma^2}=\epsilon$, with $\epsilon$ being a (small) user-defined threshold and $\rho_c$ a communication radius. 
Alternatively, other bounded continuously differentiable functions can be considered, for example
\begin{equation}
    \phi(p_i,p_j)=-0.5 \tanh{(\tau \|p_i-p_j\|_2 -\tau \rho_c)}+0.5,
\end{equation}
where $\tau$ is a user-defined parameter that allows to tune the slope of the function depending on the communication radius $\rho_c$.
The strict inequality imposed by constraint~\eqref{seq:connetivityConstr} cannot be directly implemented in a nonlinear programming (NLP) solver. A common approach is to tighten the constraint.
To this aim, let us consider an arbitrary small minimum value of the Fiedler eigenvalue $\underline{\lambda_{2}} >0$ and let us impose $\lambda_2(L(x_{(j|k)})) \geq \underline{\lambda_{2}}$ instead of constraint~\eqref{seq:connetivityConstr}.

\subsubsection{Terminal Ingredients}
Since, especially for nonlinear systems, the design of a (possibly time-varying) terminal region can be particularly challenging, in this work, we consider a terminal equality constraint corresponding to a steady-state (see~\eqref{seq:steadyState}). %Extensions to different terminal constraints, is challenging for the considered problem. 
There are many alternatives that can enlarge the domain of attraction of the approach at the cost of increased complexity in the definition of the terminal ingredients, see e.g. the work proposed by~\cite{limon2018nonlinear,ROSOLIA:2017,CARRON:2020,CHEN1998}. However, including the connectivity constraint in the computation of terminal ingredients is non trivial problem, and it is left as an open research direction.

\subsubsection{Theoretical analysis}

Before stating the results of the proposed approach, let us consider the following assumption.
% \begin{assum} \label{ass:seqsteadystate}
% For some, eventually small, $\varepsilon>0$ there exist a minimum prediction horizon $\underline{N}\in\mathbb{N}$, such that, for any steady-state $(\bar{x}_i,\bar{u}_i)$ of each agent, there exist at least a vector of input sequences $\Tilde{U}$ that generate an obstacle free trajectory allowing to preserve the connectivity while satisfying the constraints \eqref{eq:constraints} and allowing each agent to reach a steady state $(\tilde{\bar{x}}_i,\tilde{\bar{u}}_i)$ such that $l_{i,N}(C_i{\tilde{\bar{x}}_i}-r_i)\leq \max(0, l_{i,N}(C_i{\bar{x}_i}-r_i)-\varepsilon)$.
% Furthermore, we assume that at each time step, the set of admissible targets $\pazocal{R}=\{(x,u): \tilde{r}_i = C_ix_i, \ x_i = f_i(x_i,u_i), \ (x_i,u_i)\in(\mathbb{X}_i,\mathbb{U}_i), \ \forall i \in \{1, \dots, M \} \}$ is convex.
% \end{assum}
\begin{assum} \label{ass:seqsteadystate}
For the given initial condition $x_i(0)\ \forall i \in {1,\ldots,M}$, there exists a vector of input sequences $\Tilde{U}$ of length $\underline{N}\in\mathbb{N}$ that generates an obstacle free trajectory, preserves the connectivity constraints \eqref{eq:constraints}, and allows each agent to reach the reference $r_i$.
\end{assum}
Assumption~\ref{ass:seqsteadystate} implies that there exists a sequence of inputs sequences that allows the system to reach the target without violating the constraints. In case the target is not reachable, the proposed formulation allows to reach the closest reachable steady state, for more details see~\cite{limon2018nonlinear}.

The following proposition summarizes the theoretical guarantees of the proposed MPC control scheme. 
\begin{prop} \label{prop:recFeas_stab}
Let Assumptions~\ref{ass:position_invariance},~\ref{ass:cost} and ~\ref{ass:seqsteadystate} hold, and assume that the FHOCP~\eqref{eq:FHOCP} at time $k = 0$
is feasible.
Then, problem~\eqref{eq:FHOCP} is recursively feasible and, by selecting a prediction horizon $N\geq \underline{N}$, the system controlled by solving the FHOCP~\eqref{eq:FHOCP} in a receding horizon fashion, 
%the MPC control law $\kappa(x,r)$ 
converges to the reference~$r$ while satisfying the generalized connectivity condition~\eqref{eq:generalized_connectivity} and constraints~\eqref{eq:constraints} and~\eqref{eq:couplingConstr}, $\forall k\geq0$.
\end{prop}
In the following, we provide a sketch of the proof of this result.
\begin{pf}
The aforestated result comes directly from the use of the terminal equality constraint~\eqref{seq:steadyState} and classic arguments from MPC theory, see for example~\cite{rawlings2017model,limon2018nonlinear}.
The proof is divided in two parts; first we show that problem~\eqref{eq:FHOCP} is recursively feasible and then, we show that under Assumptions ~\ref{ass:position_invariance},~\ref{ass:cost} and ~\ref{ass:seqsteadystate} the system's position converges to the target. \\
\textit{Recursive feasibility:} As standard in MPC (see e.g.,~\cite{rawlings2017model}), given a feasible solution of the FHOCP~\eqref{eq:FHOCP} at time $k$ denoted as $U_i^*(k)=\{u_i^*(0|k),\dots\\ \dots,u_i^*(N-1|k)\}$ we can define the following candidate solution at time $k+1$: $\hat{U}_i(k+1)=\{u_i^*(1|k),\dots,u_i^*(N-1|k),\bar{u}_i^*\}$ leading to the candidate state trajectories $\hat{X}_i(k+1)=\{x_i^*(1|k),\dots,x_i^*(N-1|k), \bar{x}^*_i\}$, $\forall i \in \mathbb{N}_1^M$.
It is easy to show that the candidate trajectory is a feasible solution for problem~\eqref{eq:FHOCP}, thanks to the state and connectivity constraint satisfaction of the terminal equality constraint (see constraints~\eqref{seq:steadyState}~\eqref{seq:stateconstraints},~\eqref{seq:couplingstateconstraints} and ~\eqref{seq:connetivityConstr}) and thanks to the invariance of the steady-state $\bar{x}_i$ under the terminal steady-state control action $\bar{u}_i$.
\\
\textit{Convergence:} 
Under Assumptions~\ref{ass:cost} and~\ref{ass:seqsteadystate}, thanks to the results on the convergence of tracking MPC in~\cite{rawlings2017model}, the state converges to the target, i.e., $\lim \limits_{k\rightarrow \infty}\|p(k) -r \|_2\rightarrow 0$.
%Under Assumptions~\ref{ass:cost} and~\ref{ass:seqsteadystate}, in~\cite{limon2018nonlinear} is shown that by adding the additional artificial reference $\bar{r}_i$ and by introducing a convex offset cost in~\eqref{eq:CostFun} that reaches its minimum in $\bar{r}_i=r_i$, the system is able to converge to the optimal reachable steady-state. Thus, by iteratively considering Assumption~\ref{ass:seqsteadystate}, we obtain that for $k$ that tends to infinity, $l_{i,N}(\bar{r}_i - r_i)\rightarrow0$ and thus $\lim \limits_{k\rightarrow \infty}\|p(k) -r \|_2\rightarrow 0$.
\end{pf}
\begin{rem}
Note that the results presented in Proposition~\ref{prop:recFeas_stab} remain valid also in the presence of a time-varying reference $r$, thanks to the offset cost in~\eqref{eq:CostFun}, as shown in~\cite{limon2018nonlinear}.
\end{rem}
\section{Distributed SQP-based Solution}
\label{sec:sqp}
The FHOCP~\eqref{eq:FHOCP} is an NLP due to the nonlinearities introduced by the system dynamics~\eqref{seq:sysdyn} and the nonlinear constraint~\eqref{seq:connetivityConstr}. %A general-purpose NLP solver can be used to solve the optimization problem at hand. However, this requires the adoption of a central unit that collects the state of the agents and runs the optimization, which is impractical in many multi-agent scenarios. 
Addressing the problem in a centralized way would require communicating the state of all agents to a single unit, and then computing the solution and sending back the optimizer in real-time. This may be impractical in many cases, due to delays and limited computation capabilities. To avoid the use of a central-unit, we propose an SQP-based solution, for more details see~\cite{boggs1995sequential}, that is amenable to distributed optimization. 

In order to obtain the SQP formulation, we first write the FHOCP~\eqref{eq:FHOCP} as
\begin{subequations}\label{eq:NLP}
    \begin{align}
        \min \limits_{z} \;\;
         & f_{NLP}(z)               \\
        \text{s.t.:} \ \ \ \
        % DYNAMICS
         & g(z)=0,        \\
         & h(z)\leq 0,  \\
         & -\lambda_2(L(z))+\underline{\lambda_{2}}\leq 0 \label{seq:connectConstr_NLP},
    \end{align}
\end{subequations}
where $z$ is the optimization variable that includes both state and input. 
It can be noticed that we have highlighted constraint~\eqref{seq:connetivityConstr} in~\eqref{seq:connectConstr_NLP}. In this section, we aim to show how the gradient of the connectivity constraint and the related term of the Hessian of the Lagrangian, which will be better defined in the following paragraph, can be analytically computed.

Sequential quadratic programming computes a solution to~\eqref{eq:NLP}
by iteratively solving quadratic programs that locally approximate the original NLP at the current solution estimate~$\hat{z}_{j}$. At each iteration, the solution $\Delta z$ of the quadratic subproblem
\begin{subequations}\label{eq:SQP}
    \begin{align}
        \min \limits_{\Delta z} \;\;
          \nabla &f_{NLP}(\hat{z}_{j})^T\Delta z + \frac{1}{2}\Delta z^T H\Delta z \label{seq:SQP_cost}\\
        \text{s.t.:} \ \ \ \
        % DYNAMICS
         & g(\hat{z}_{j})+\nabla g(\hat{z}_{j})\Delta z=0,  \\
         & h(\hat{z}_{j})+\nabla h(\hat{z}_{j}) \Delta z\leq 0,  \\
         & -\lambda_2(L(\hat{z}_{j}))+\underline{\lambda_{2}}-\left. {\frac{\partial \lambda_2(L(\hat{z}_{j}))}{\partial z}}\right|_{\hat{z}_{j}} \Delta z\leq 0 \label{seq:SQP_connectConstr},
    \end{align}
\end{subequations}
is used to update the current solution estimate, where $\nabla f_{NLP}$, $\nabla g$ and $\nabla h$ denote the gradients of $f_{NLP}$, $g$ and $h$, respectively. In the above optimization problem, the matrix $H$ denotes the Hessian of the Lagrangian~$\pazocal{L}$ of~\eqref{eq:NLP}, where~$\pazocal{L}$ is defined as
\begin{equation*}
    \pazocal{L}(z,\mu^g,\mu^h,\mu^{\lambda}) = f_{NLP}(z)+\mu^{g^T} g(z)+\mu^{h^T} h(z)-\mu^{\lambda}\lambda_2(L(z)),
\end{equation*}
and $\mu^g,\mu^h,$ and $\mu^{\lambda}$ are the Lagrange multipliers.
The Hessian, in general, can be approximated, e.g. with Broyden–Fletcher-Goldfarb–Shanno (BFGS) algorithm, \\ see~\cite{wright1999numerical}, or exactly computed. Also the gradients required to solve~\eqref{eq:SQP}, can be analytically or numerically computed. In this paper, we provide a result that shows how the gradient and the Hessian of the Lagrangian term related to the connectivity constraint~\eqref{seq:connectConstr_NLP} can be analytically computed by exploiting Theorem~\ref{th:theremEig}. In fact, according to~\eqref{eq:dsigma}, we can write:
\begin{figure*}[ht] 
\centering
\subfloat[][\label{subfig:Conn}]
		{\includegraphics[width=.45\textwidth]{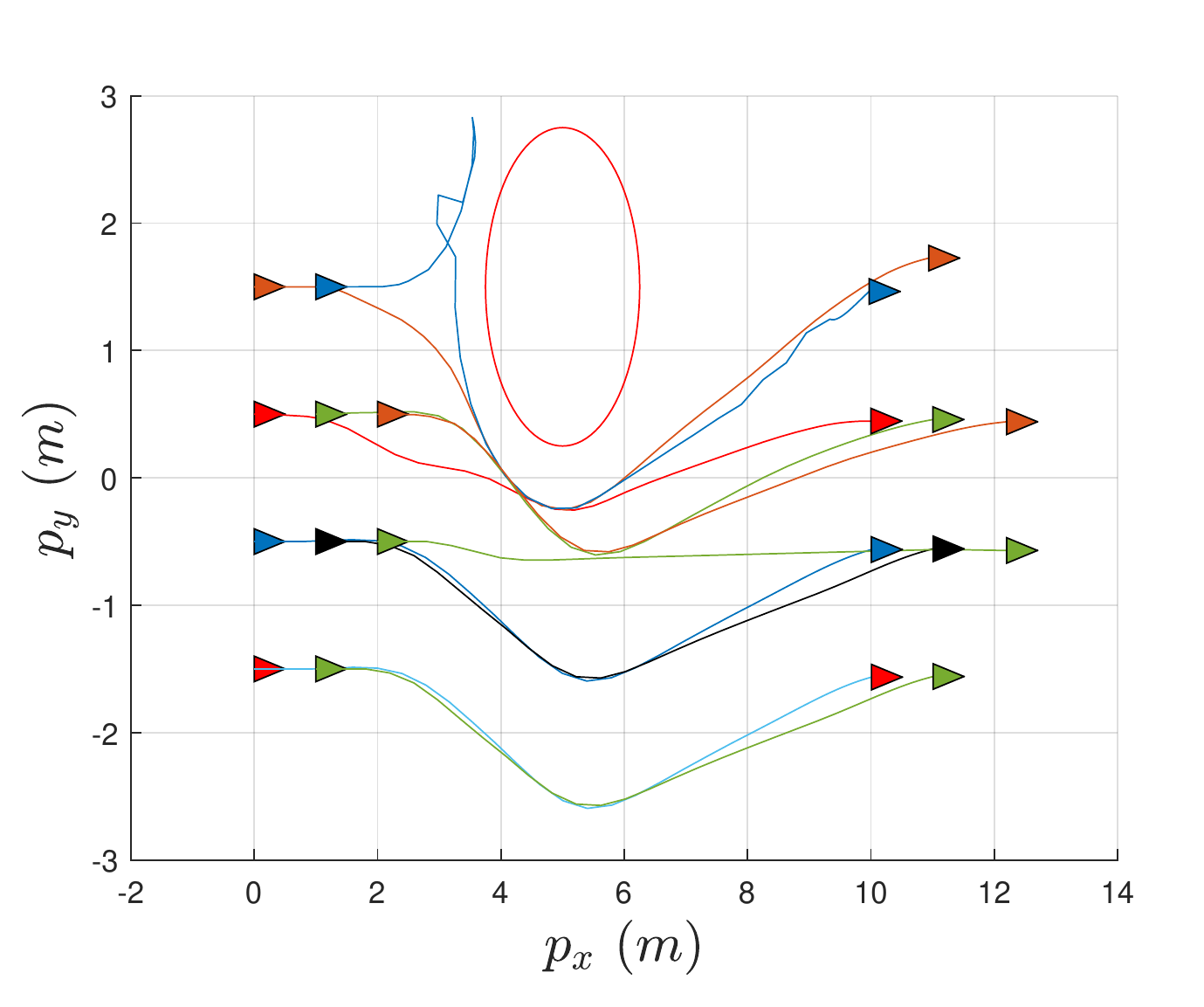}}\quad
\subfloat[][\label{subfig:Nocon}]
		{\includegraphics[width=.45\textwidth]{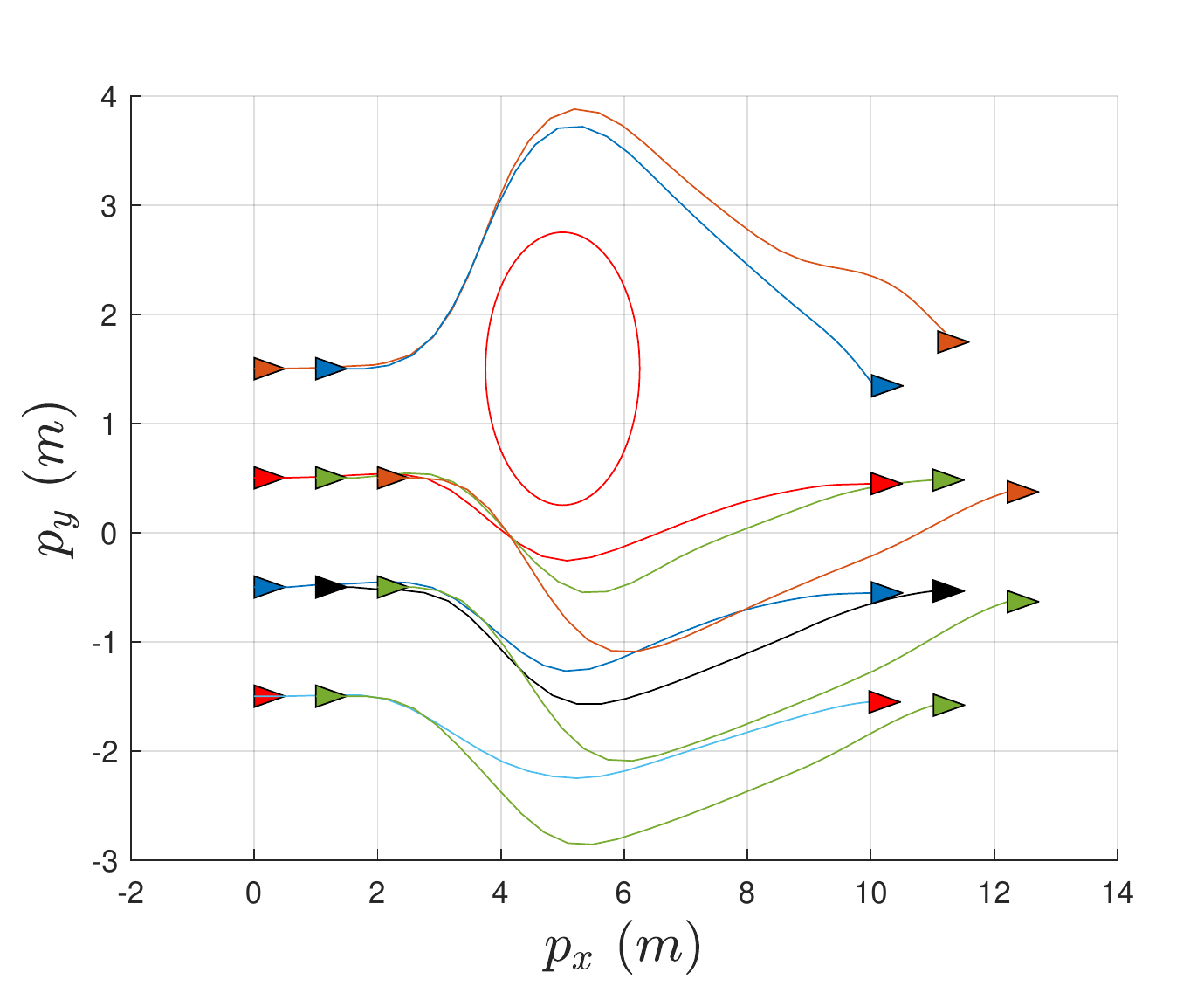}}\quad
\caption{Simulation of the proposed approach with ten agents. Closed-loop trajectories obtained by solving the FHOCP~\eqref{eq:ex_FHOCP} with (Fig.~\ref{subfig:Conn}) and without (Fig.~\ref{subfig:Nocon}) the connectivity constraint~\eqref{seq:example_connConstr}.}
\label{fig:Simulation}
\end{figure*}
\begin{equation} \label{eq:dlambda2}
    \frac{\partial \lambda_2 }{\partial x_i} = v_2^T \frac{\partial L}{\partial x_i} v_2 = \sum_{j\in \pazocal{N}_i}{\frac{\partial A_{ij}}{\partial x_i} (v_{2}^i-v_{2}^j)},
\end{equation}
where~$v_{2}^i$ is the~$i$-th entry of the eigenvector~$v$ associated with the second smallest eigenvalue~$\lambda_2$ of the Laplacian~$L$. \\
Furthermore, by differentiating a second time~\eqref{eq:dsigma}, we obtain that 
% by considering~\eqref{eq:dsigma}, the second differential of the eigenvalue function is:
% \begin{equation} \label{eq:ddlambda2}
%     \text{d}^2\lambda_2 = \text{d}v_2^T(\text{d}L)v_2+v_2^T(\text{dd}L)v_2+v_2^T(\text{d}L)\text{d}v_2,
% \end{equation}
% where d$v_2$ is defined as in~\eqref{eq:dv}.
% \\ 
the second derivative of the Fiedler eigenvalue with respect to the state $x$ can be written as:
\begin{equation} \label{eq:ddlambda2}
    \frac{\partial^2\lambda_2}{\partial x_ix_j}=
    \left[
        \frac{\partial v_2}{\partial x_j}
        \right]^T
    \left[
        \frac{\partial L}{\partial x_i}v_2
        \right]+
    v_2^T \left(
    \frac{\partial^2 L}{\partial x_i x_j}
    \right)
    v_2+
    v_2^T
    \frac{\partial L}{\partial x_i}
    \frac{\partial v_2}{\partial x_j},
\end{equation}
where $\frac{\partial v_2}{\partial x_i}$ is derived by~\eqref{eq:dv}.
%Note that, the second derivative of constraint~\eqref{seq:connectConstr_NLP} required in~\eqref{seq:SQP_cost}, can be analytically computed with equation~\eqref{eq:ddlambda2} or with suitable toolboxes for automatic differentiation (see e.g. \cite{andersson2019casadi}), while the first derivative required in~\eqref{seq:SQP_connectConstr} comes directly from~\eqref{eq:dlambda2}. 
In the following, we show that the QP-subproblem~\eqref{eq:SQP} maintains the graph structure inherited from the communication network, and thus can be solved using distributed optimization techniques, such as the Alternating Direction Method of Multipliers (ADMM) (see~\cite{BOYD_ADMM:2011}). 
In the original problem~\eqref{eq:FHOCP}, the coupling among neighbouring agents is imposed by constraints~\eqref{seq:couplingstateconstraints} and~\eqref{seq:connetivityConstr}, i.e., the coupling state constraint and the connectivity constraint. By definition, the first only couples neighbouring agents, and thus also its linearization in~\eqref{eq:SQP} preserves the same sparsity pattern imposed by the communication network. Regarding the connectivity constraint, it is possible to write~\eqref{seq:SQP_connectConstr} such that it only depends on local information. In fact, the smallest eigenvalue~$\lambda_2$ of the Laplacian matrix $L$ and the associated eigenvector~$v_2$ can be computed in a distributed fashion by following the approach proposed in~\cite{YANG_CONNECTIVITY:2010}, which makes use of a modified power-iteration method. Finally, the Jacobian~\eqref{eq:dlambda2} and Hessian~\eqref{eq:ddlambda2} of the second smallest eigenvalue of the Laplacian only depend on local information. In particular for the Jacobian, it is possible to see that the summation in~\eqref{eq:dlambda2} runs only over the neighbouring agents thus requiring only local information. The Hessian computation in~\eqref{eq:ddlambda2} comprises three terms. The computation of the first and third terms require only local information as they involve a multiplication with the derivative of the Laplacian with respect to the state of agent $i$. The same argument holds for the second term as it involves the second derivative of the Laplacian with respect to the state of agent $i$ and $j$.

To conclude, the SQP-based algorithm to solve optimization problem~\eqref{eq:FHOCP} is summarized in Algorithm~\ref{alg:sqp}.
\setlength{\algomargin}{1.3em}
\begin{algorithm} [h]
    \caption{SQP-based algorithm}
    \label{alg:sqp}
    Initialize $\hat{z}_0,\mu^{g}_0,\mu^{h}_0,\mu^{\lambda}_0$, and $j=0$\\ 
    \While{\text{ termination criterion not met }}{
        Compute $\lambda_2$,  $\frac{\partial \lambda_2}{\partial z}$, and $\frac{\partial^2 \lambda_2}{\partial  z^2}$ at $\hat{z}_j$\\
        Compute Hessian $H$ of $\pazocal{L}$ at $(\hat{z}_{j},\mu^{g}_{j},\mu^{h}_{j},\mu^{\lambda}_{j})$\\
        Compute $\nabla f_{NLP}$, $\nabla g$, $\nabla h$ at $\hat{z}_j$ \\
        Solve~\eqref{eq:SQP}  \label{sqp:SolveQP}\\
        Get the optimal multipliers of~\eqref{eq:SQP}: $\mu^g_{qp}$, $\mu^h_{qp}$, $\mu^{\lambda}_{qp}$\\
        Set $\mu^g_{j+1} = \mu^g_{qp}$, $\mu^h_{j+1} = \mu^h_{qp}$, $\mu^{\lambda}_{j+1}=\mu^{\lambda}_{qp}$ \\
        Set $\hat{z}_{j+1} \ = \hat{z}_{j} + \alpha_j \Delta z$\\
        Set $j = j + 1$
    } 
\end{algorithm}
\\
In Algorithm \ref{alg:sqp}, the parameter $\alpha_j$ is determined in order to produce a sufficient decrease in a merit function that combines the objective with measures of constraint violation, see \cite{wright1999numerical}, to avoid steps that reduces the objective function but increases the violation of the constraints.
\begin{rem}
In general, the Hessian of the Lagrangian $L$ may not be positive definite. However, to guarantee the convergence of the SQP-based algorithm, it is necessary that the obtained approximation of $H$ is positive definite. When this is not the case, there exists various approaches to find a positive definite approximation, see for example~\cite{Verschueren2017convexification}.
\end{rem}
% \begin{rem}
% Note that the QP defined in~\eqref{eq:SQP} is not guaranteed to be recursive feasible. Its feasibility can be guaranteed by properly relaxing some constraints, or by enforcing suitable techniques beyond the scope of this paper, see for example~\cite{lawrence2001computationally,zanelliinexact}.
% \end{rem}

%%%%%%%%%%%%%%%%%%%%%%%%%%%%%%%%%%%%%%%
\section{Numerical example}
\label{sec:numerical}
\begin{figure}
	\centering
	\includegraphics[width=\columnwidth]{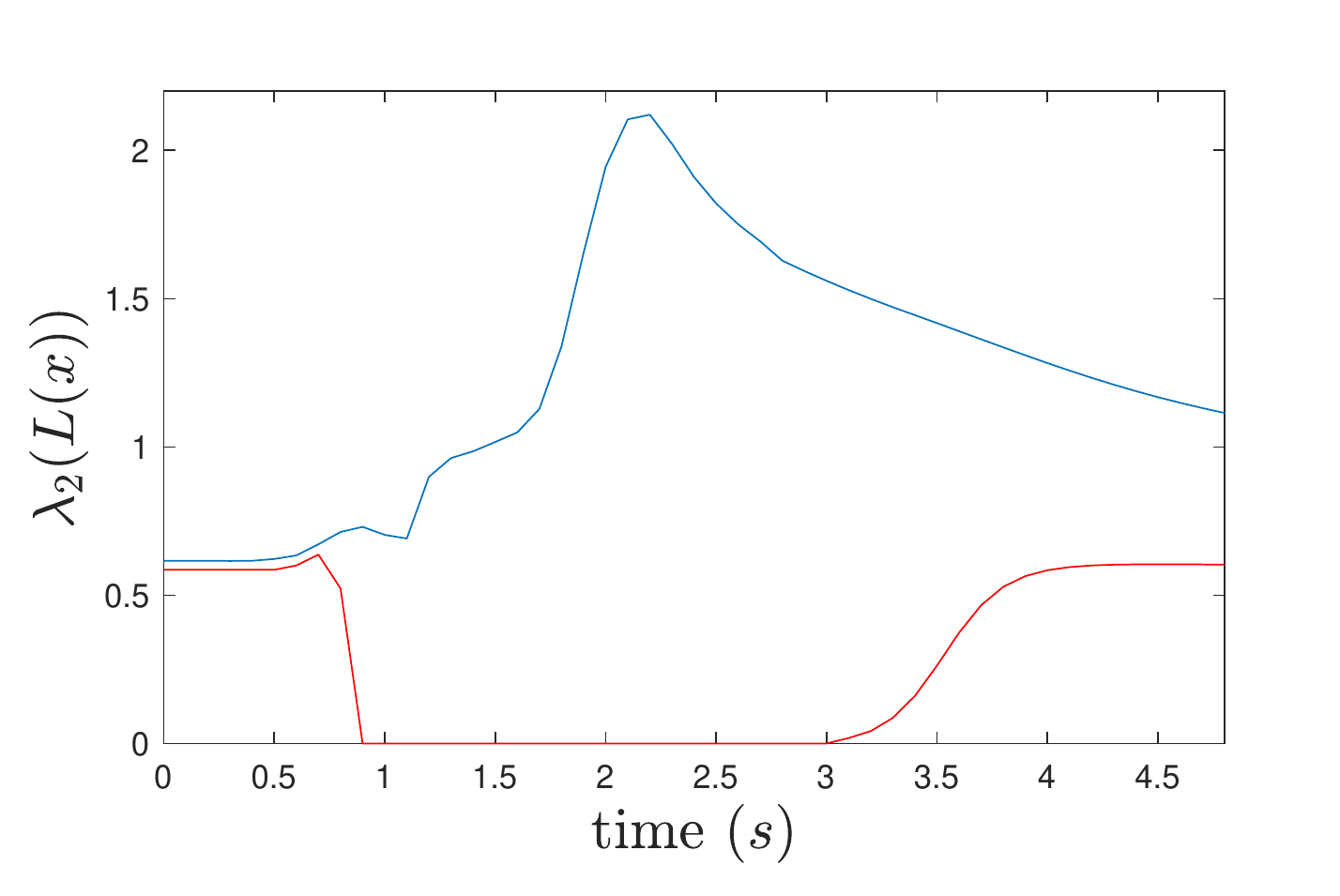}
	\caption{Value of $\lambda_2(L(x))$ obtained during the simulations when constraint~\eqref{seq:example_connConstr} is implemented (blue line) and when it is not included (red line). }	
	\label{fig:comp_lambda}
\end{figure}

We provide two numerical examples to show the effectiveness of the proposed approach. The simulations are performed in MATLAB$^\circledR$ on a Quad-Core Intel Core i7 (2.8 GHz, 16 GB) machine under MS Windows, using CasADI to compute the derivative needed to define the QP~\eqref{eq:SQP} and \verb|quadprog| to solve each iteration of the SQP.
For both examples, we consider a swarm of vehicles, each one described by the following non-linear time-invariant dynamics:
\begin{align}
    \label{eq:model_example}
     x_i(k+1) &= f(x(k),u(k)), \nonumber\\
    \begin{split}
      \begin{bmatrix}
      p_{x_i}(k+1)\\
      p_{y_i}(k+1)\\
      \theta_i(k+1)\\
      \gamma_i (k+1)
      \end{bmatrix}  & = \left(
    \begin{matrix}
    p_{x_i}(k)+T_s \cos{(\theta_i(k))}v_i(k)
    \\
    p_{y_i}(k)+T_s \sin{(\theta_i(k))}v_i(k)
    \\
    \theta_i(k) + T_s \frac{v_i(k)}{L} \tan{(\gamma_i(k))}
    \\
    \gamma_i(k)+T_s \delta_i(k)
    \end{matrix} \right),\\
     p_i(k) &= Cx_i(k), \nonumber
    \end{split}
\end{align}
where $T_s=0.1$ is the sampling time,~$L=0.005$ is the wheelbase of the agent,~$p_{xi}$,~$p_{yi}$ represent the position of the agent,~$\theta_i$ the yaw angle and~$\gamma_i$ the steering angle.  The input of the system is represented by the commanded velocity $v_i$ and steering rate~$\delta_i$.
The yaw angle of each agent is constrained to be within the set~$[-\pi, \pi]$, while the steering angle~$\gamma_i\in[-\frac{\pi}{2}, \frac{\pi}{2}]$, defining the set of state constraints~$\mathbb{X}_i$.
The velocity of each agent is instead limited in the range~$[-4, 4]$~$m/s$ and the steering rate~$|\delta_i|\leq1$, defining the set of input constraints~$\mathbb{U}_i$.
Each agent has a circular communication set~$\cC_i(p_i)=\{q:\|p_i - q\|_2 \leq \rho_c \}$ with a communication radius~$\rho_c=2$ m. To define the adjacency matrix, we consider the following function~$\phi(p_i,p_j)=-0.5\tanh(10\|p_i - q\|_2-10\rho_c)+0.5$.
Furthermore, to avoid collisions between different agents, we consider an obstacle avoidance constraint in the FHOCP, by setting as forbidden area a ball with %In particular, we consider the agents described by a ball with
diameter $\rho_a = 1$ m around the agent position.
Thus, we impose the collision avoidance constraint~$\|p_i - p_j\|_2 \geq \rho_a$. 
\\
As a first example, we present a scenario where a swarm of~$M=10$ agents has to cross the workspace, while avoiding collisions with other agents and with a circular obstacle centered at~$\xi_o$ and of radius~$\rho_o$ , see Fig.~\ref{fig:Simulation}.
Thus, we consider the following FHOCP
\begin{subequations}\label{eq:ex_FHOCP}
    \begin{align}
        &\min \limits_{U,\bar{x},\bar{u},\bar{r}} \;\;
          \sum_{i=1}^{M} J_i(x,U,\bar{x}_i,\bar{u}_i,\bar{r}_i,r_i) \nonumber \\
        % DYNAMICS
         &\text{subject to: } \nonumber \\
         & x_{i}(0|k)=x_i(k),  \\
         & x_{i}(j+1|k)=f(x_{i}(k|t),u_{i}(k|t)), && \forall j \in \mathbb{N}_0^{N-1}, \\
         & x_{i}(j|k)\in\mathbb{X}_i, && \forall j \in \mathbb{N}_0^{N}, \\
         & u_{i}(j|k)\in\mathbb{U}_i, && \forall j \in \mathbb{N}_0^{N-1}, \\
         & \|p_{i}(j|k)-\xi_o \|_2\geq(\frac{\rho_a}{2}+\frac{\rho_a}{2}), && \forall j \in \mathbb{N}_0^{N}, \\
         & ||p_{i}(j|k)-p_{l}(j|k) ||_2\geq\rho_a,  \forall l \in \pazocal{N}_i, && \forall j \in \mathbb{N}_0^{N},   \\
         & \bar{x}_i = f(\bar{x}_i,\bar{u}_i),\\
         & x_{i}(N|k)=\bar{x}_i, \ \  \bar{u}_i \in \mathbb{U}_i\\
         & \bar{r}_i = C \bar{x}_i, \\
         & \lambda_2(L(x))\geq 0.1, \label{seq:example_connConstr}\\
         & \forall i \in \mathbb{N}_1^M,
    \end{align}
\end{subequations}
the considered tracking cost function is
\begin{multline}
    J_i(x,U,\bar{x}_i,\bar{u}_i,\bar{r}_i,r_i)=\sum_{j=1}^{N-1}|x_{i}(j|k)-\bar{x}_i|_Q+\\
    +|u_{i}(j|k)-\bar{u}_i|_R  +|\bar{r}_i-r_i|_{Q_{2}},
\end{multline}
where $Q=I_4$, $R=0.5 \ I_2$, $Q_2=5 \ I_4$ and the selected prediction horizon is $N=10$.
The obtained NLP, is solved with Algorithm \ref{alg:sqp}. Fig.~\ref{fig:Simulation} shows the closed-loop trajectories of the agents applying the control law obtained by solving~\eqref{eq:ex_FHOCP}, either when constraint~\eqref{seq:example_connConstr} is implemented (Fig.~\ref{subfig:Conn}) or without (Fig.~\ref{subfig:Nocon}).
%. On the left (Fig.~\ref{subfig:Conn}) shows the trajectories when constraint~\eqref{seq:example_connConstr} is implemented, in contrast, on the right (Fig.~\ref{subfig:Nocon}) is shown the behaviour of the system without the constraint~\eqref{seq:example_connConstr}. 
The value of the Fiedler eigenvalue~$\lambda_2(L(x))$ during the two simulations is shown in Fig.~\ref{fig:comp_lambda}. The red line shows the behaviour of the eigenvalue when the connectivity constraint~\eqref{seq:example_connConstr} is not implemented. As it can be noticed, when the red agent in Fig.~\ref{subfig:Nocon} circumnavigates the obstacle, 
%the communication between all the agents is lost and 
the eigenvalue drops to $0$ meaning that the graph is disconnected and the network has no connectivity. The blue line, instead, depicts the value of $\lambda_2(L(x))$ when constraint~\eqref{seq:example_connConstr} is implemented. As it can be noticed in Fig.~\ref{subfig:Conn}, in this case, the red agent reduces its velocity and circumnavigates the obstacle on the other side to keep communication with the rest of the swarm.\\
\begin{figure*}
	\centering
	\includegraphics[width=\textwidth]{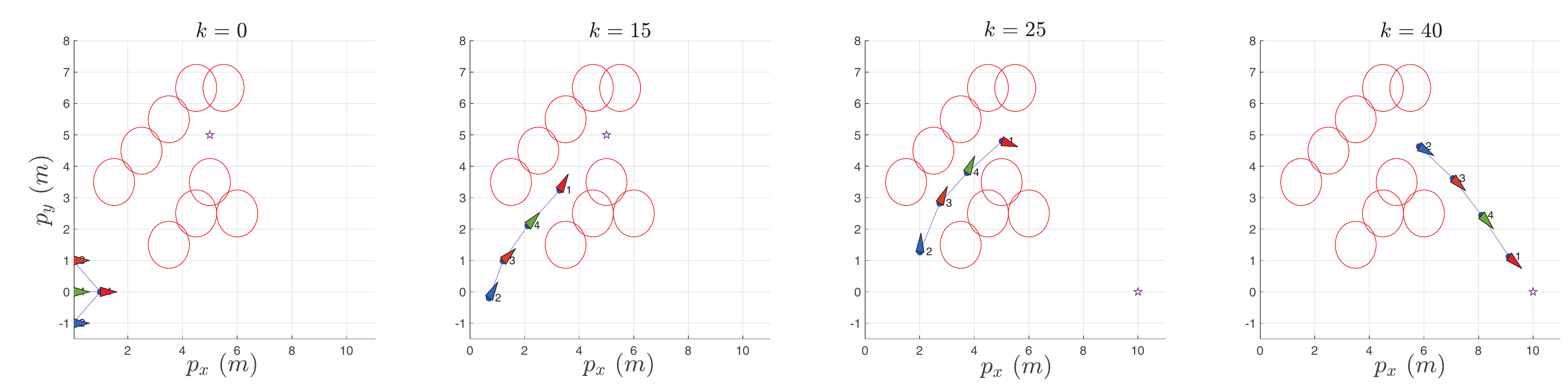}
	\caption{Simulation of the proposed approach in a leader-follower setup at iteration $k=0,15,25,40$. Blue lines represent the communication graph, red circles are the considered obstacles while the purple star is the position reference $r_1(k)$.}	
	\label{fig:leader_ex}
\end{figure*}
The second scenario, shown in Fig.~\ref{fig:leader_ex}, represents a leader-follower approach where $M=4$ agents are considered.
The leader, depicted as a red triangle in Fig.~\ref{fig:leader_ex}, follows a time-varying reference $r_1$ that drives the leader through a series of obstacles. To this aim, the cost function of problem~\eqref{eq:ex_FHOCP} is replaced with the following one:
\begin{multline}
    J(x_1,U,\bar{x}_1,\bar{u}_1,\bar{r}_1,r_1) = \sum_{j=1}^{N-1}{|x_{1}(j|k)-\bar{x}_{1}|_Q}+\\+|u_{1}(j|k)-\bar{u}_1|_R 
    +|\bar{r}_{1}-r_1|_{Q_{2}}.
\end{multline}
Fig.~\ref{fig:leader_ex} shows the simulation at time $k=0,15, 25$ and $40$. The communication graph is highlighted in blue. It can be noticed that the connectivity constraint~\eqref{seq:example_connConstr} forces the follower agents to follow the leader to maintain the communication.
%%%%%%%%%%%%%%%%%%%%%%%%%%%%%%%%%%%%%%%
\section{Conclusions}
\label{sec:conclusions}
This paper presented a multi-agent model predictive controller able to enforce network connectivity by imposing a constraint on the connectivity eigenvalue while guaranteeing input and state constraint satisfaction. To implement the proposed approach based on only local computation, we propose a SQP reformulation of the original nonlinear optimal control problem. Numerical simulations show the effectiveness of the approach.

\bibliography{bibliography}             % bib file to produce the bibliography
\end{document}